%% file: MAIN.tex
\documentclass[sigconf]{acmart}

\usepackage{xcolor}
\usepackage{fontawesome}
\usepackage{multirow}
\usepackage{enumitem}
\usepackage{titlesec}
\usepackage{breakurl}
\usepackage{hyperref}
\hypersetup{colorlinks,allcolors=black}

\setlist{nosep}
\titlespacing{\section}{0pt}{6pt}{2pt}
\newcommand{\sectopic}[1]{\par\noindent{\textit{\bfseries #1}}}

\AtBeginDocument{%
  }

\copyrightyear{2024}
\acmYear{2024}
\setcopyright{rightsretained}
\acmConference[ICSE-Companion '24]{2024 IEEE/ACM 46th International Conference on Software Engineering: Companion Proceedings}{April 14--20, 2024}{Lisbon, Portugal}
\acmBooktitle{2024 IEEE/ACM 46th International Conference on Software Engineering: Companion Proceedings (ICSE-Companion '24), April 14--20, 2024, Lisbon, Portugal}
\acmDOI{10.1145/3639478.3639810}
\acmISBN{979-8-4007-0502-1/24/04}

\begin{document}

\title{Generating User Experience Based on Personas with AI Assistants}

\author{Yutan Huang}
\affiliation{
  \institution{HumaniSE Lab, Faculty of Information Technology, Monash University}
  \city{Clayton}
  \state{Victoria}
  \country{Australia}
}
\email{yutan.huang1@monash.edu}

% \affiliation{%

% \affiliation{%

\begin{abstract}
 
Traditional UX development methodologies focus on developing ``one size fits all" solutions and  lack the flexibility to cater to diverse user needs. In response, a growing interest has arisen in developing more dynamic UX frameworks. However, existing approaches often cannot personalise user experiences and adapt to user feedback in real-time. Therefore, my research introduces a novel approach of combining Large Language Models and personas, to address these limitations. The research is structured around three areas: (1) a critical review of existing adaptive UX practices and the potential for their automation; (2) an investigation into the role and effectiveness of personas in enhancing UX adaptability; and (3) the proposal of a theoretical framework that leverages LLM capabilities to create more dynamic and responsive UX designs and guidelines.
\end{abstract}
\maketitle



\input{Files/introduction}

\input{Files/Related_work}

\input{Files/Research_Questions}

\input{Files/Solution_Approach}

\input{Files/Discussion}
\bibliographystyle{ACM-Reference-Format}
\balance
\bibliography{paper}

\end{document}

%% file: Files/introduction.tex
\section{Introduction}
\vspace{0.75em}

User Interface (UI) and User Experience (UX) are integral components in software engineering (SE) that serve to bridge the gap between human requirements and system functionalities. UI and UX aim to optimise the interaction between the computer and the human via the interface to ensure ease of use and intuitiveness. A well-implemented UI/UX not only diminishes the cognitive load on the user but also reduces the time and effort required for users to understand and navigate through a system~\cite{kashfi2014models}. Hence, properly designed UI/UX significantly affects system efficiency, user satisfaction, and overall performance~\cite{stone2005user}.
In the rapidly advancing technological landscape, users' desire for customised options and personalised experiences has surged, emphasising the importance of customisable and adaptive UX~\cite{yang2020measuring}. In addition, there is a growing recognition of the necessity for human-centric requirements that cater to individuals with specific needs, such as those with disabilities or diverse backgrounds~\cite{grundy2020humanise}. \emph{Customizable} UX allows users to control and tailor the design based on their preferences. It represents an important step toward user-centric interfaces but often fails to deliver a truly personalised experience~\cite{hui2015enhancing,liu2003adaptive}. \emph{Adaptive} UX goes beyond customisation, employing the ability to understand user behaviours, preferences and context~\cite{main2020guru}. Consequently, the system proactively alters the elements of UI to serve users better, e.g., visual appearance, typography, colour schemes, iconography and interactive elements like buttons, forms, and navigation menus~\cite{kashfi2017integrating}. While the idea of a truly adaptive system seems appealing, its practical implementation is challenging due to the diverse needs of users. Additionally, manually designing such a system is laborious, compounded by the need to maintain consistency due to business requirements, e.g., branding and aesthetics. 

Personas are often used in the field of UX as archetypical user profiles to inform designers about specific user behaviours, needs and goals from the system~\cite{salminen2022use}. Their strength lies in providing a clear, focused understanding of end-users, especially when direct access to human beneficiaries is limited, enabling designers to make informed decisions. The recent advances in artificial intelligence (AI) techniques offer great potential for adaptive UI and addressing the challenges mentioned above via automation. Large Language Models (LLMs) are the recent successors in the area of AI techniques that have shown considerable promise in automating different SE tasks, e.g., code generation~\cite{le2022coderl}, requirements management~\cite{arora2023advancing}, test generation~\cite{nguyen2023generative}, and persona generation~\cite{zhang2023personagen,zhu2023personalized}. LLMs, trained on vast amounts of data, are excellent candidates for generating adaptive designs due to their ability to understand context, infer user intentions, and generate coherent responses~\cite{fan2023large}. This PhD research intends to explore the potential of LLMs combined with rich personas, which are more comprehensive and detailed than standard personas, to develop adaptive UX for diverse users. Specifically, I aim to create an adaptive UX framework that tailors user interfaces according to individual preferences and needs, focusing on the design, adapting and leveraging personas (and user requirements).

Next, I discuss the related work on adaptive UX and the use of personas (Section~\ref{sec:related work}), and the research plan with research questions (RQs) (Section~\ref{sec:Research questions}). This PhD project is in the early stages; hence, in Section~\ref{sec:Solution Approach}, I discuss the proposed approach and research directions.

%% file: Files/Related_work.tex
\section{Related work}~\label{sec:related work}
\vspace{-0.5em}

Adaptive UI/UX design uses a model-based approach as well as an AI-based approach~\cite{stige2023artificial,todi2021adapting}. The model-based approach involves the creation of adaptive designs using architectural models. These models consist of one or multiple layers of architecture that process multimodal data to generate adaptive UXs~\cite{lehmann20103}. This approach primarily focuses on enhancing UX features such as layout, content, and modality, however, while it achieves diversification by leveraging different models, it often lacks the invaluable input of user feedback and iterative refinement derived from legacy systems~\cite{akiki2016engineering}. Additionally, the methodology for runtime feature selection is often underdeveloped in this approach, which limits its ability to adapt to changing user needs and preferences~\cite{hussain2018model}. This model-based approach seeks to create variations in UX but may fall short in addressing real-time user interactions and feedback~\cite{hussain2018model}.

In contrast, the AI-based approach has gained prominence in recent years, capitalizing on the capabilities of AI to generate both text and graphics. Researchers have employed AI tools such as Sketch2Code, MetaMorph, and ChatGPT to dynamically generate UIs based on user interactions and requirements~\cite{potluri2019ai,wen2023droidbot}. The use of AI in adaptive UX design introduces a range of possibilities. Yang et al. identified four key channels through which AI augments the value of adaptive UX: self-inferences, world inferences, optimal inferences, and utility inferences. These channels represent AI's ability to provide users with self-understanding, contextual understanding, optimal solutions, and utility-based responses, significantly enriching the user experience~\cite{yang2018mapping}. These four channels serve as foundational concepts for adaptive UX generation with AI and are essential for guiding designers to create more personalized and user-centric interfaces~\cite{ekvall2023integrating}. Despite the potential of AI-based approaches, it's becoming increasingly evident that solutions utilizing Large Language Models (LLMs) are at the forefront of this technology's application. These LLMs, which are now among the most commonly implemented forms of AI, heavily rely on the quality of prompts provided to them~\cite{naveed2023comprehensive}. In the context of user experience (UX) design, these prompts' precision and relevance directly impact the outcomes' quality, as demonstrated in recent studies~\cite{liu2023jailbreaking}. Effective prompt engineering is a critical aspect of AI-driven adaptive UX requirements, and it is an area that requires careful consideration and refinement~\cite{arora2023advancing}.
The model-based and AI-based approaches in adaptive UX design have illustrated diverse possibilities. However, it's important to note that these approaches commonly lack rigorous evaluation and iterative feedback from users and designers, forming a significant gap in the existing research landscape. This review provides the context for understanding the need for our research, which aims to address these limitations and enhance the field of adaptive UX design by constructing an intelligent User interface that uses ML techniques with a framework to guide experts through the process of creating adaptive UI with user experience.

%% file: Files/Research_Questions.tex
\section{Research Plan}~\label{sec:Research questions}
\vspace{-0.2em}

The main research aim of this PhD research is to develop a framework for generating adaptive UX using LLMs and personas structured in the following steps (guided by the research questions mentioned under each step).

\sectopic{Foundational Understanding}: How is adaptive UX defined and understood in the current literature? Which UX fragments can be adapted and generated automatically? %How does it differ from traditional UX methodologies?
\sectopic{Role of Personas in Adaptive UX}:  What are the critical elements within personas that lend themselves to the creation of adaptive UX? Are there gaps or limitations in current persona models that could hinder the development of adaptive UX designs?
\sectopic{Role of LLMs in Adaptive UX}: To what degree can LLMs contribute to the development of adaptive UX? How do LLMs interpret and utilise persona information to generate UX designs? Which prompting techniques in LLMs yield the best adaptive UX results?
\sectopic{Framework Development and Evaluation}: Do users and practitioners find the adaptive UX generated by our framework useful? What are the challenges when leveraging LLMs for adaptive UX?

%% file: Files/Solution_Approach.tex
\section{Solution Approach}~\label{sec:Solution Approach}
\sectopic{Foundational Understanding - } \textit{Systematic Literature Review and UX experiment}: My foundational understanding begins with a systematic literature review on adaptive UI/UX, exploring definitions, methods, and applications in academic and professional contexts to identify aspects of UX that have been automated previously. \textcolor{black}{Concurrently, I will conduct experiments to create UI automatically using LLMs, with insights from the literature, to validate my findings and identify potential UI fragments that can be adapted easily (e.g., interface designs, colours, buttons).} This will establish a foundation for developing an informed adaptive UI/UX framework.
\sectopic{Role of Personas in Adaptive UI/UX - } \textit{Expert Insight and Model Comparison}: To figure out the important parts of personas that help create adaptive UI and find any shortcomings in current persona representations, I will find key persona elements related to adaptive UI in practice by interviewing experienced UX designers. I will then compare different representations of persona contents and prioritise what is important to include in a persona for adaptive UX generation. \textcolor{black}{The comparative analysis and interviews in parallel will help refine persona representations and triangulate our findings.}
\sectopic{Role of LLMs in Adaptive UI/UX - }\textit{Exploring LLM's Capability in Adaptive UI Creation}: I plan to carry out a set of experiments revolving around prompt engineering, an example would be using GPT-model-based LLMs and feeding them user preference and background information with personas. These experiments can examine the effectiveness of LLMs in generating user-tailored designs.
\sectopic{Framework Development and Evaluation - } \textit{Evolving UI/UX Framework through User and Practitioner Feedback}: I aim to develop a UX framework based on LLMs to guide adaptive UX creation. This framework will be dynamic, evolving through iterative enhancements for robustness and effective adaptive UX design. Leveraging LLM capabilities, I seek to establish a foundational, adaptable tool for UX development.

\textit{Assessment and Refinement of the UI/UX Framework through User-Centric Feedback}: \textcolor{black}{The evaluation of adaptive UI design and UI/UX framework will involve engaging users and experts to interact with and test the developed UIs by using them as a daily routine and provide users with tasks to complete.}\textcolor{black}{Their feedback will inform the integration of prompt engineering into our framework, and enhancing a smooth transition from design-time to run-time approach.} 
    
    %Post-feedback, we will create guidelines for using our framework and continuously refine it to ensure reliability in creating adaptive UX with LLMs. 

%% file: Files/Discussion.tex
\section{Conclusion}~\label{sec:Discussion}
\vspace{0.1em}

In conclusion, the research aims to address a gap in adaptive UX design by integrating LLMs and personas, striking a balance between technical capabilities and a human-centric approach. The challenge lies in aligning LLMs' technical prowess with the nuanced insights of personas. The expected impact includes enhanced adaptability and personalization in UX designs, and setting new standards in UX methodology. 

\section*{Acknowledgement}
\vspace{5pt}

Yutan is supported by ARC Laureate
Fellowship FL190100035.